\documentstyle[aps,prl,multicol,epsfig]{revtex}

\begin{document}

\draft

\title{Entangling macroscopic oscillators exploiting radiation pressure}

\author{
Stefano Mancini$^{1,3}$, 
Vittorio Giovannetti$^{2}$,
David Vitali$^{3}$
and Paolo Tombesi$^{3}$}

\address{
$^{1}$INFM, Dipartimento di Fisica,
Universit\`a di Milano,
Via Celoria 16,
I-20133 Milano, Italy
\\
$^{2}$Research Laboratory of Electronics,
MIT - Cambridge, MA 02139, USA
\\
$^{3}$INFM, Dipartimento di Matematica e Fisica,
Universit\`a di Camerino,
I-62032 Camerino, Italy} 

\date{\today}

\maketitle

\begin{abstract}
It is shown that radiation pressure can be profitably used to
entangle {\it macroscopic} oscillators like movable mirrors, using
present technology. We prove a new sufficient criterion for
entanglement and show that the 
achievable entanglement is robust
against thermal noise. Its signature can be revealed using 
common optomechanical readout apparatus.
\end{abstract}


\begin{multicols}{2}

The fundamental role of {\it entanglement} \cite{SCH}
in quantum mechanics has been
re-emphasized in recent years \cite{ENT}. 
In this context, an important point is to asses whether this
peculiarity of the quantum world, i.e., the entanglement, could be
applicable to macroscopic bodies and, moreover, measurable.

Literature focused on methods to 
prepare atoms in entangled states already exists
\cite{SAK}. The recent experiment generating entanglement
of two gas samples \cite{JUL} is a striking achievement.
Then, a real challenge is to devise the 
possibility of applying similar 
arguments to macroscopic, massive oscillators.
Here we propose an experiment, which could be realized with present
technologies, to show that it is possible to entangle massive oscillators 
exploiting the radiation pressure force.

It is indeed usually believed that, being a
superposition of states, the entanglement between massive, 
macroscopic,
objects is practically impossible to detect because of the 
fast diagonalization of the
system's density matrix due to the coupling with the environment
\cite{ZUR}.
On the contrary, by using a new sufficient criterion
for entanglement, we shall derive the parameter region for which 
two massive, movable, cavity mirrors can be entangled by the 
radiation pressure exerted by a cavity mode, and we shall show how to 
measure the degree of entanglement.
Beside foundation interest, the ability to place such oscillators 
in entangled states may even result useful in applications, as
in high precision measurements \cite{HOLL}. 

To be concrete, as a specific model we consider two end 
mirrors of an optical cavity, which 
can both oscillate under the 
effect of radiation pressure force.
Cavities with one movable mirror have already been 
studied \cite{VAR}, 
and a wide class 
of quantum states resulting from optomechanical coupling was 
proposed \cite{NCS}.
Furthermore, due to recent technological developments in 
optomechanics, this area is now
becoming experimentally accessible \cite{EXP}.

\begin{figure}[t]
\centerline{\epsfig{figure=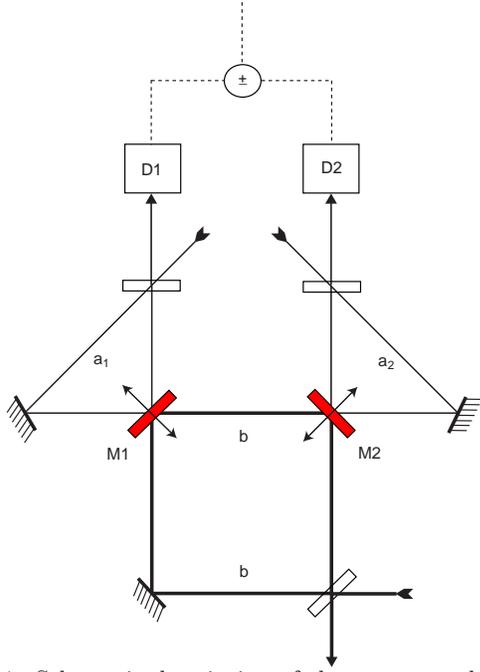,width=2.5in}}
\caption{\narrowtext 
Schematic description of the system under study.
For the sake of simplicity the oscillating mirrors $M1$ and
$M2$ are assumed to be identical.
An intense light field $b$ (entangler) couples the 
moving mirrors.
Their tiny movements (indicated by the arrows)
are then detected through  
the meter modes $a_1$ and $a_2$ which are subjected 
to homodyne measurement at $D1$ and $D2$.
Finally, the two output currents are combined 
to get center of mass or relative mirrors coordinate.
}
\label{fig1}
\end{figure}

As pointed out in Ref. \cite{BRAG}, under the assumption 
that the measurement time is either less or of the order of 
the mechanical relaxation 
time, it is possible to consider a macroscopic 
oscillator, i.e., a movable mirror in our case, 
as  a quantum oscillator. 
Then, for not too high oscillation frequency, with 
respect to the inverse round trip times of
photons within the cavities, we can write the 
Hamiltonian of the system sketched in 
Figure \ref{fig1} as
\begin{eqnarray}\label{H}
{\cal H}&=&\sum_{i=1}^{2}\hbar\omega_{a}
a_{i}^{\dag}a_{i}
+\hbar\omega_{b} b^{\dag}b
+\hbar\Omega\sum_{i=1}^{2}\left(
\frac{p_{i}^{2}}{2}+\frac{q_{i}^{2}}{2}\right)
\nonumber\\
&&-\hbar g a^{\dag}_{1} a_{1} q_{1}
+\hbar g a^{\dag}_{2} a_{2} q_{2}
+\hbar G b^{\dag} b \left(q_{1}-q_{2}\right)
\nonumber\\
&&+i\hbar\sqrt{\gamma_a}\sum_{i=1}^{2}
\left(\alpha^{in}e^{-i\omega_{a0}t}a_i^{\dag}
-\alpha^{in\,*}e^{i\omega_{a0}t}a_i\right)
\nonumber\\
&&+i\hbar\sqrt{\gamma_b}\left(\beta^{in}e^{-i\omega_{b0}t}b^{\dag}
-\beta^{in\,*}e^{i\omega_{b0}t}b\right)\,,
\end{eqnarray}
where $a_{i}$, $a_{i}^{\dag}$
are the destruction and creation operators of the electromagnetic fields
corresponding to the meters mode and $\omega_a$ their
frequency (assumed equal for simplicity).
Instead, $b$, $b^{\dag}$ are those of the {\it entangler} mode
(the use of this terminology will become clear in the following) and
$\omega_b$ its frequency. Finally, $q_{i}$ and
$p_{i}$ are the dimensionless 
position and momentum
operators of the mirrors $M_i$, both oscillating at frequency $\Omega$,
and having mass $m$. 
The first row of equation (\ref{H}) simply represents
the free Hamiltonian, whereas the second
represents the effect of the radiation 
pressure force which causes the 
instantaneous displacement of the mirrors \cite{LAW}.
The coupling constants are
$g={\tilde g}/\sqrt{m\Omega}$ and $G={\tilde G}/\sqrt{m\Omega}$ 
where ${\tilde g}$, ${\tilde G}$ are related to the
cavity mode frequencies, to
the equilibrium length of the cavities, and to the reflection angles
\cite{VAR,LAW}. The last two rows represent the driving fields action
in the usual rotating wave approximation.
We assume that both 
meters ($a_1$, $a_2$) 
are driven at frequency $\omega_{a0}$,
while the entangler mode $b$ is 
driven at frequency $\omega_{b0}$;
$\alpha^{in}$, $\beta^{in}$ are the classical 
fields characterizing the input 
laser powers
$P_a^{in}= \hbar \omega_{a0} |\alpha^{in}|^{2}$,
$P_b^{in}= \hbar \omega_{b0} |\beta^{in}|^{2}$, and
$\gamma_{a}$, $\gamma_b$ are the cavity linewidths.

By considering, the unitary evolution
of the two mirrors and the entangler, neglecting the meters modes and 
the driving terms in Eq. (\ref{H}), 
it can be easily checked that 
the formers become entangled once a von Neumann
projection onto the $b$-mode quadrature is 
performed (this is why we named such mode entangler).
A detailed analysis of the problem, however, must include 
photon losses, the
thermal noise on the mirrors, and the measurement backaction.
It means that the interaction of all optical modes with their respective 
reservoirs and the effect of thermal fluctuations on the two mirrors, not 
considered in Hamiltonian (\ref{H}), must be added to this equation. This 
can be accomplished in the 
standard way \cite{milwal,VIT}. 
The resulting Hamiltonian   
gives rise to nonlinear 
Langevin equations whose linearization around the 
steady state leads to
\begin{equation}\label{LINEQS}
\begin{array}{l}
\dot{a}_j = i\Delta_a a_j + (-)^{(j+1)} i g \alpha q_j
-\frac{\gamma_a}{2} a_j+\sqrt{\gamma_a} a^{in}_j  \,,  
\\ \\
\dot{b} = i\Delta_b b-i G\beta (q_1-q_2) -  
\frac{\gamma_b}{2} b+\sqrt{\gamma_b} b^{in}\,,
\\ \\
\dot{q}_j=\Omega p_j \,,
\\ \\
\dot{p}_j=-\Omega q_j+(-)^{(j+1)}g\alpha(a_j+a^{\dag}_j)
\\ \\ 
\hspace{0.8 in}
+(-)^j G (\beta^* b+\beta b^{\dag})-\Gamma p_j+\xi_j \,,
\end{array} 
\end{equation}
where $j=1,2$, and all the operators now represent 
small fluctuations
around steady state values.
These are
\begin{eqnarray}
\begin{array}{l}
\langle q_j \rangle_{ss}
= (-)^{j}[G|\beta|^2-g|\alpha|^2]/\Omega,
\\ \\  
\langle p_j \rangle_{ss} = 0,
\\ \\
\alpha \equiv  \langle a_j \rangle_{ss} 
=\sqrt{\gamma_{a}} \alpha^{in}/
[\gamma_a/2 -i\Delta_a],
\\ \\
\beta \equiv \langle b \rangle_{ss}
=\sqrt{\gamma_b} \beta^{in}
[\gamma_b/2-i\Delta_b].
\end{array}
\end{eqnarray}
Moreover, $\Delta_a\equiv\omega_{a0}-\omega_a
+g\langle q_1 \rangle_{ss}$,
$\Delta_b\equiv\omega_{b0}-\omega_b
-G(\langle q_1 \rangle_{ss}-\langle q_2 \rangle_{ss})$,
are the radiation phase shifts due to the detuning and
to the stationary displacement of the mirrors. Both radiation fields
used as meters ($a_1$, $a_2$) 
are damped through output fixed
mirrors at the same rate $\gamma_a$,
while the entangler mode $b$ is damped at rate $\gamma_b$.
Furthermore, $\Gamma$ is the mechanical damping rate for the 
mirrors Brownian motion.
Without loss of generality, we choose $\alpha$ real 
and $\Delta_a=0$. 
The operators $a^{in}_{j}(t)$ and $b^{in}(t)$ 
represent the vacuum (white) noise operators 
at the cavity inputs.
The noise operator for the
quantum Brownian motion 
of the mirrors is $\xi_j(t)$.
The non-vanishing noise correlations are
\begin{eqnarray}\label{NOISE}
&&\langle a^{in}_j(t) a^{in\,\dag}_k(t') 
\rangle = 
\delta(t-t')\,\delta_{j,k}\,,\quad j,k=1,2\,,
\nonumber\\
&&\langle b^{in}(t) b^{in\,\dag}(t') 
\rangle = 
\delta(t-t')\,,
\\
&&\langle {\xi_j}(t) {\xi_k}(t') \rangle =
\delta_{j,k}\int \, d\omega \,  
\frac{\Gamma\omega}{2\Omega}
\frac{
\left[\coth\left(
\hbar\omega/2k_BT\right)-1 \right]} 
{e^{i\omega(t-t')}}
\,,\nonumber
\end{eqnarray}
where $k_B$ is the Boltzmann constant and $T$ the 
equilibrium temperature (the two mirrors are considered
in equilibrium with their respective bath at the same 
temperature). Notice that the used approach for the Brownian
motion is quantum mechanical consistent at every temperature
\cite{VIT}.

The unitary evolution under the linearized Hamiltonian
leading to system of Eqs. (\ref{LINEQS}) gives 
entanglement, as in the non-linearized case discussed above.
Hence, the main task is to see whether such quantum 
correlations are visible or blurred by noisy 
effects.
To accomplish this task, we first solve the 
system (\ref{LINEQS}) in the frequency domain
by introducing the pseudo Fourier transform
${\cal O}(\omega)=\tau^{-1/2} 
\int_{-\tau/2}^{\tau/2}
\, dt\, e^{i\omega t} {\cal O}(t)$ for each operator ${\cal O}$, 
where $\tau$ is the measurement time 
assumed to be large compared to all the relevant timescales of the 
system dynamics.

Let us now consider the measured current at each meter
output. The boundary relations for the meters radiation fields \cite{GAR},
i.e., $a^{out}_j = \sqrt{\gamma_a}a_j-a^{in}_j$,
yield the phase quadratures $Y_j=-i(a_j-a^{\dag}_j)$ 
at the output, namely
\begin{equation}\label{YOUT}
Y^{out}_j(\omega)=\frac{2g\alpha\sqrt{\gamma_a}}
{\gamma_a/2-i\omega}q_j(\omega)
+\frac{\gamma_a/2+i\omega}{\gamma_a/2-i\omega}
Y^{in}_j(\omega)\,.
\end{equation} 
Thus, the measurement of the output quadrature $Y^{out}_j$, 
in the detection box $D_j$,
indirectly gives the mirror position $q_j$.
More precisely, in homodyne detections, the positive 
and negative frequency components of the 
quadrature being measured are combined through
a proper modulation, in order to achieve 
the measurement of a hermitian operator \cite{GAR}.
Then, it would be 
possible to indirectly measure either $[q_j(\omega)+q_j(-\omega)]$
or $i[ q_j(-\omega)-q_j(\omega)] $,
which implies the possibility to
measure position or momentum
for each macroscopic oscillator.

These measurements can be used to establish when the two oscillating
cavity mirrors are entangled. Sufficient criteria for entanglement
of continuous variable systems already exist \cite{DUAN,SIMON}, but 
here we shall introduce a {\em new} 
sufficient inseparability criterion, involving the {\em 
product} of variances of continuous observables:

{\em Theorem}. If we define
$u=q_{1}+q_{2}$ and $v=p_{1}-p_{2}$,
then, for any separable
quantum state $\rho $, one has
\begin{equation}  \label{e3}
\left\langle \left( \Delta u\right) ^{2}\right\rangle 
\left\langle \left( \Delta v\right) ^{2}\right\rangle 
\geq |\langle[q_1,p_1]\rangle|^2 \,.
\end{equation}
See the appendix for the proof.

This theorem allows us to establish a connection with Refs.~\cite{REID}, 
which showed that  
when the inequality
\begin{equation}\label{INFER}
\left\langle \left( \Delta u\right) ^{2}\right\rangle 
\left\langle \left( \Delta v\right) ^{2}\right\rangle 
< \frac{1}{4}|\langle[q_1,p_1]\rangle|^2\,,
\end{equation}
is satisfied, an EPR-like paradox arises\cite{EPR}, based on the 
inconsistency between quantum mechanics and local realism.
Notice that the sufficient condition for inseparability of 
Eq.~(\ref{e3}) is weaker than condition (\ref{INFER}), but this is not 
surprising, since entangled states are only a necessary condition for the 
realization of an EPR-like paradox.

The theorem (\ref{e3}) can then be used to establish the conditions 
under which the two massive oscillators are entangled. In fact, 
defining the hermitian operator 
${\cal R}_{\{{\cal O}\}}(\omega)=[{\cal O}(\omega)+{\cal O}(-\omega)]/2$
for any operator ${\cal O}(\omega)$ in the frequency domain,
and using Eq.~(\ref{e3}), 
we define the degree of entanglement ${\cal E}(\omega)$ as
\begin{equation}\label{EDEF}
{\cal E}(\omega)=\frac{
\langle {\cal R}^2_{\{u\}}(\omega) \rangle
\,
\langle {\cal R}^2_{\{v\}}(\omega) \rangle}
{ \left|
\langle \left[{\cal R}_{q_1}(\omega),{\cal R}_{p_1}(\omega)\right]
\rangle\right|^2}\,,
\end{equation}
(we use the fact that 
$\langle u\rangle = \langle v \rangle =0$ in our case)
which is a marker of entanglement whenever ${\cal E}(\omega) < 1$.
If, moreover, it goes below $1/4$, that indicates the presence
of EPR correlations.

To calculate the function ${\cal E}(\omega)$ 
we evaluate the correlations 
\begin{equation}
\langle {\cal O}(\omega){\cal O}(\pm\omega) \rangle
=\int_{-\tau/2}^{\tau/2} \, \frac{dt}{\tau}\,  
\int_{-\infty}^{\infty} \, dt' \,
e^{i\omega t'} 
\langle {\cal O}(t){\cal O}(t'\mp t) \rangle\,,
\end{equation}
and use the solutions of (\ref{LINEQS}) and  
correlations (\ref{NOISE}) in the frequency domain.
In doing that,
we require $G>g$ and $P_b^{in}>P_a^{in}$ because a strong interaction 
between mirrors and entangler is desirable. The strength
of the system-meter interaction, instead, has to guarantee 
only a 
sufficient measurement gain. This condition, 
by referring to Eq.(\ref{YOUT}), corresponds to  
$g^2\alpha^2\gg(\gamma_a^2/4+\omega^2)/4$.

\begin{figure}[t]
\centerline{\epsfig{figure=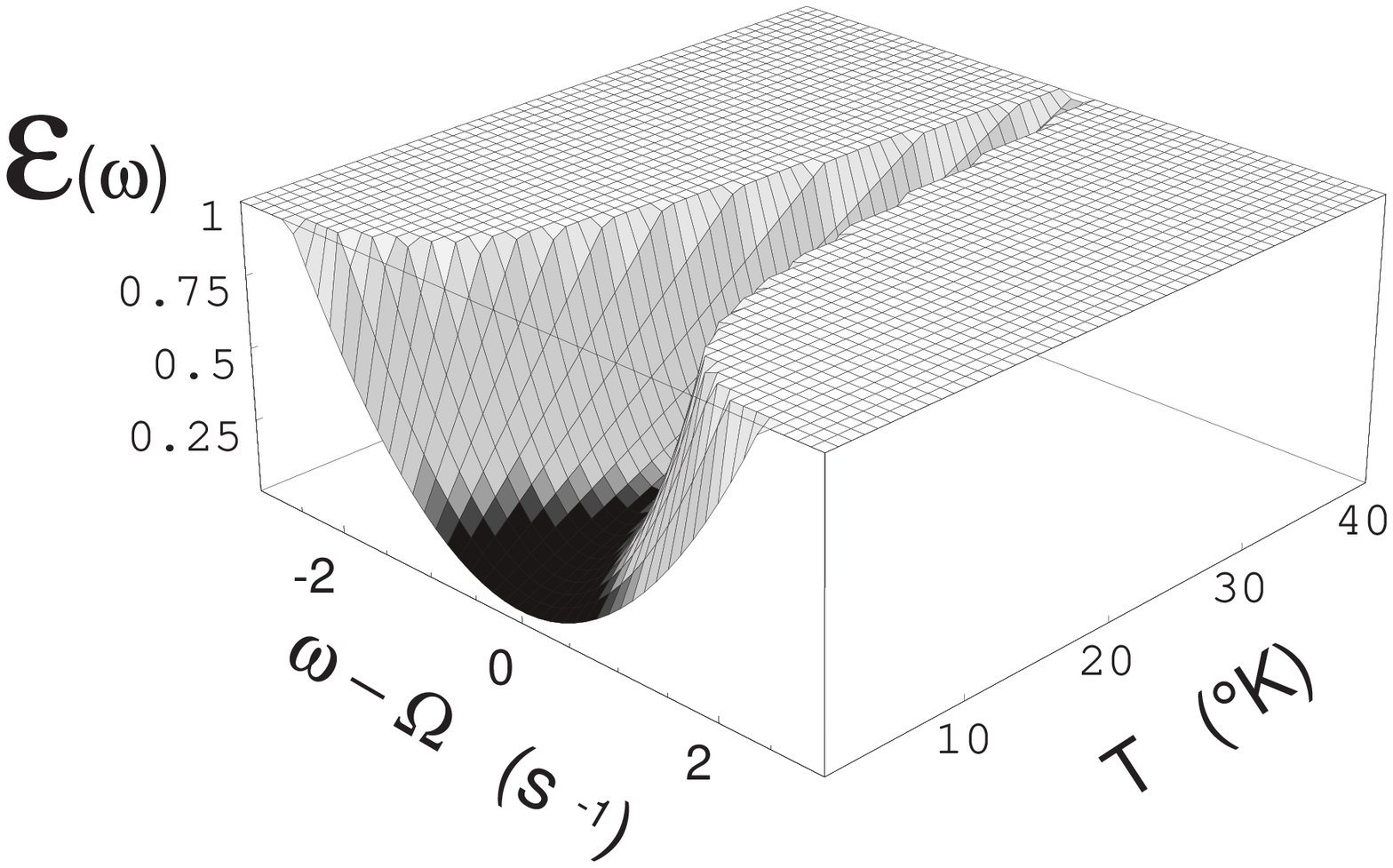,width=3.5in}}
\caption{\narrowtext 
Degree of entanglement ${\cal E}$ as function of 
frequency $\omega$ and temperature $T$.
The plot has been cut at ${\cal E}=1$, and 
the part of surface for which
$0\le {\cal E}<1/4$ is black coloured.
The value of parameters are:
$\gamma_a=\gamma_b=\Delta_b=10^{5}$ ${\rm s}^{-1}$;
$P^{in}_a=5\times 10^{-4}$ $W$;
$P^{in}_b=5\times 10^{-3}$ $W$;
$\Omega=10^{5}$ ${\rm s}^{-1}$;
$m=10^{-5}$ ${\rm Kg}$;
$\Gamma=1$ ${\rm s}^{-1}$;
$g=0.5$ ${\rm s}^{-1}$;
$G=5$ ${\rm s}^{-1}$.
With these parameters the cavity lengths
are $\approx 10^{-2}$ ${\rm m}$ for the $b$ mode
and $\approx 10^{-1}$ ${\rm m}$ for the $a$ modes.
}
\label{fig2}
\end{figure}

In Fig.\ref{fig2} we show the behavior of the 
degree of entanglement (\ref{EDEF}) as function of frequency
and temperature for massive oscillators with $m=10^{-5}$
${\rm Kg}$ and $\Omega=10^5$ ${\rm s}^{-1}$. The maximum entanglement 
is always 
obtained at the frequency $\Omega$ of the oscillating mirrors
where the mechanical response is maximum.
The useful bandwidth becomes narrower and tends to 
disappear as the temperature increases.
Nevertheless, a large amount of entanglement is
available at reasonable temperatures 
e.g. $4$ ${}^{\circ}K$.
It means to have purely quantum effects 
at macroscopic scale notwithstanding $k_BT\gg\hbar\Omega$.
It is also worth noting that the values of parameters 
here employed are essentially those already used in experiments 
\cite{EXP}. Moreover, considering single mode oscillators, as we have 
done here, is not a restrictive assumption because the various 
internal and external oscillating modes of the mirrors have different 
oscillation frequencies and they can be easily distinguished and 
addressed when measurements are performed in the frequency domain. 
Other promising candidates for the realization of entanglement
between two massive objects are given by mesoscopic resonators,
such as microfabricated cantilevers \cite{CLELAND}.

As can be evicted from Fig.\ref{fig2}, at low temperatures,
${\cal E}(\Omega)$ lies below the limit $1/4$. 
Thus, the studied system also provides an example of 
{\it macroscopic}
EPR correlations, though with
the experimental set-up of Ref.\cite{EXP},
a further condition, 
concerning the spatial
separation between the two systems, is required to test 
the paradox \cite{EPR}.
However, other possible set-ups
could be devised permitting even such test. 
Demonstration of entanglement is
instead much less demanding.

In conclusion, we have exploited the ponderomotive force to 
entangle macroscopic oscillators.
Reliable conditions to achieve this goal are 
established by also accounting
for a measurement of the degree of entanglement.
The obtained results appears quite robust
against the thermal noise and
could be challenging tested with current 
technologies opening new perspectives
towards the use of Quantum Mechanics
in macroscopic world. 
Moreover, the possibility to prepare entangled 
state at the macroscopic level may prove to be useful for high 
precision and metrology applications. For example, it 
is possible to see that a scheme similar to that 
of Fig.~1 can be used to improve the detection of weak forces 
\cite{stefa}.

{\em Appendix}

We prove the sufficient criterion for inseparability
for the pair
of continuous variable operators
$u =|a|q_{1}+\frac{1}{a}q_{2}$ and
$v =|a|p_{1}-\frac{1}{a}p_{2}$,
where $a$ is an arbitrary (nonzero) real number.
Assuming $\rho = \sum_{i}w_{i}\;\rho _{i1}\otimes \rho _{i2}$
and using the same first steps of the proof of Ref.~\cite{DUAN}, we have
\begin{eqnarray}
&&\left\langle \left( \Delta u\right) ^{2}\right\rangle 
\left\langle \left( \Delta v\right) ^{2}\right\rangle 
=
\left\{\sum _{i}w_{i}\left( a^{2}\left\langle \left(
\Delta q_{1}\right) ^{2}\right\rangle _{i} \right. \right.
\nonumber \\
&&\left. \left. +\frac{1}{a^{2}}
\left\langle \left( \Delta q_{2}\right) ^{2}\right\rangle
_{i}\right) +
\sum_{i}w_{i}\left\langle u
\right\rangle _{i}^{2}-\left( \sum_{i}
w_{i}\left\langle u\right\rangle _{i}\right) ^{2}\right\}
\nonumber \\
&&\times
\left\{\sum _{i}w_{i}\left( a^{2}\left\langle \left(
\Delta p_{1}\right) ^{2}\right\rangle _{i}+\frac{1}{a^{2}}
\left\langle \left( \Delta p_{2}\right) ^{2}\right\rangle
_{i}\right) \right.\nonumber \\
&&+ \left.
\sum_{i}w_{i}\left\langle v
\right\rangle _{i}^{2}-\left( \sum_{i}
w_{i}\left\langle v\right\rangle _{i}\right) ^{2}\right\},
\end{eqnarray}
where the symbol $\left\langle \cdots \right\rangle _{i}$ denotes
average over the product density operator $\rho _{i1}\otimes \rho _{i2}$. By
applying the Cauchy-Schwarz inequality $\left( 
\sum_{i}w_{i}\right) \left( \sum_{i}
w_{i}\left\langle u\right\rangle _{i}^{2}\right) \geq \left( 
\sum_{i}w_{i}\left| \left\langle u\right\rangle _{i}\right|
\right) ^{2},$ we can rewrite
\begin{eqnarray}
&&\left\langle \left( \Delta u\right) ^{2}\right\rangle 
\left\langle \left( \Delta v\right) ^{2}\right\rangle 
\geq \nonumber \\
&& \left\{\sum _{i}w_{i}\left( a^{2}\left\langle \left(
\Delta q_{1}\right) ^{2}\right\rangle _{i}+\frac{1}{a^{2}}
\left\langle \left( \Delta q_{2}\right) ^{2}\right\rangle
_{i}\right) \right\}
\nonumber \\
&&\times
\left\{\sum _{i}w_{i}\left( a^{2}\left\langle \left(
\Delta p_{1}\right) ^{2}\right\rangle _{i}+\frac{1}{a^{2}}
\left\langle \left( \Delta p_{2}\right) ^{2}\right\rangle
_{i}\right) \right\}.
\end{eqnarray} 
Then using the fact that $\alpha^{2}+\beta^{2}\geq 2\alpha \beta$, we 
have  
\begin{eqnarray}
&&\left\langle \left( \Delta u\right) ^{2}\right\rangle 
\left\langle \left( \Delta v\right) ^{2}\right\rangle 
\geq 
4\left\{\sum _{i}w_{i}\sqrt{ \left\langle \left(
\Delta q_{1}\right) ^{2}\right\rangle _{i}
\left\langle \left( \Delta q_{2}\right) ^{2}\right\rangle
_{i}} \right\}
\nonumber \\
&&\times
\left\{\sum _{i}w_{i}\sqrt{ \left\langle \left(
\Delta p_{1}\right) ^{2}\right\rangle _{i}
\left\langle \left( \Delta p_{2}\right) ^{2}\right\rangle
_{i}} \right\}.
\end{eqnarray} 
We then use again the Cauchy-Schwartz inequality and get
\begin{eqnarray}
&&\left\langle \left( \Delta u\right) ^{2}\right\rangle
\left\langle \left( \Delta v\right) ^{2}\right\rangle
\geq 
4\left(\sum _{i}w_{i}\left[ \left\langle \left(
\Delta q_{1}\right) ^{2}\right\rangle _{i}
\left\langle \left( \Delta q_{2}\right) ^{2}\right\rangle
_{i}  \right. \right. \nonumber \\
&& \left. \left. \times \left\langle \left(
\Delta p_{1}\right) ^{2}\right\rangle _{i}
\left\langle \left( \Delta p_{2}\right) ^{2}\right\rangle
_{i} \right]^{1/4}\right)^{2},
\end{eqnarray} 
which gives the final inequality of Eq.~(\ref{e3})
when the Heisenberg uncertainty principle is applied.
  
\section*{Acknowledgements}

The authors gratefully acknowledge Mario Rasetti,
for critical reading of the manuscript.

\end{multicols}

\end{document}